\documentclass[twocolumn]{article}
\usepackage{xcolor}
\usepackage{wrapfig}
\usepackage{graphicx}
\usepackage[nolist]{acronym}
\usepackage{authblk} 
\usepackage{booktabs}
\usepackage{float}
\usepackage{hyperref}
\usepackage[margin=0.75in, left=0.45in, right=0.45in]{geometry} 

\AtBeginDocument{%
  \providecommand\BibTeX{{%
    \normalfont B\kern-0.5em{\scshape i\kern-0.25em b}\kern-0.8em\TeX}}}

\title{Poster: Developing an O-RAN Security Test Lab}

\author[1]{Sotiris Michaelides}
\author[2]{David Rupprecht}
\author[1]{Katharina Khols}

\affil[1]{Radboud University, Nijmegen, Netherlands}
\affil[2]{Radix Security, Bochum, Germany}

\date{} 

\begin{document}

\maketitle

\begin{abstract}
\ac{oran} is a new architectural approach, having been proposed only a few years ago, and it is an expansion of the current \ac{ngran} of 5G. \ac{oran} aims to break this closed RAN market that is controlled by a handful of vendors, by implementing open interfaces between the different \ac{ran} components, and by introducing modern technologies to the \ac{ran} like machine learning, virtualization, and disaggregation. However, the architectural design of ORAN was recently causing concerns and debates about its security, which is considered one of its major drawbacks. Several theoretical risk analyses related to \ac{oran} have been conducted, but to the best of our knowledge, not even a single practical one has been performed yet. In this poster, we discuss and propose a way for a minimal, future-proof deployment of an \ac{oran} 5G network, able to accommodate various hands-on security analyses for its different elements.
\end{abstract}






\section{Introduction}

The cellular infrastructure comprises three major components: the \ac{ran}, \ac{cn}, and \ac{ue}. \ac{ue}, typically a phone with a Sim card, is connected to the \ac{cn}, where all of the network functions reside. \ac{ran} is responsible to link the two ends, by providing wireless connectivity to the \ac{ue} and by being physically connected to the \ac{cn}. It is traditionally composed of a \ac{bbu} which implements the protocols of the network layer stack and an \ac{ru} responsible for Radio Frequencies. Their communication is based on a Point-to-Point protocol, that each vendor may implement according to his preferences, which limits the flexibility and scalability of \ac{ran}. Over the 4G era, RAN evolved with the BBU split into the \ac{cu} and \ac{du}, and several architectures were proposed, that reduced the costs but remained vendor-oriented, i.e vendor-specific implementations for the different \ac{ran} components.

\ac{oran} is a new approach, promoted by the O-RAN Alliance~\cite{ORAN}, that introduces modern concepts to the architecture such as virtualization, \ac{ml}, and disaggregation. Dissagretation is a method that separates the \ac{sw} from the underlying \ac{hw}, while Virtualisation uses \ac{sw} to create instances of virtual environments that run on the same \ac{hw}. These modern tools enable the deployment of agile, flexible, and intelligent RANs. 

However, the new concepts of ORAN result in new types of attacks not previously seen in RAN. Three new risk areas associated with \ac{oran} are architecture openness, cloud and virtualization, and machine learning. These attack vectors are being added to the current 5G architectural threats that are already linked to the \ac{ran}. A study by the German Federal Office for Information Security concluded that ORAN's components are vulnerable and exposed and lack basic security mechanisms \cite{5GRAN_ANALYSIS}. Other theoretical analyses have shown similar results \cite{abdalla2022generation}\cite{liyanage2022open}\cite{MIMRAN2022102890}. 

While a theoretical identify potential vulnerabilities and risks in a system before they can be exploited by attackers.
A practical analyses allows for evaluating the feasibility and the consequences of each attack in a real-world environment. To the best of our knowledge, no practical security analysis of the ORAN architecture exists. This poster proposes the deployment of a basic \ac{oran} lab for such a security analysis, at minimum cost, able to accommodate several practical risk analyses on the different ORAN components and interfaces. Our contribution is thereby not only showing the current architecture of our lab but also discussing the challenges when setting it up.

\section{DEPLOYING THE TEST LAB}

\textit{A. Requirements}\\ According to Köpsell et. al. virtualization, \ac{ml}, and architectural openness are the new threat areas related to \ac{oran}~\cite{5GRAN_ANALYSIS}. Therefore we decided that we limit our practical security analysis to the architectural openness and to \ac{ml} parts of ORAN. 

For architectural openness, the open-front-haul interface between the \ac{odu} and \ac{oru}, is the most promising interface to practically analyze. This interface uses the \ac{ecpri} protocol and the new 7.2x split, defined by the O-RAN alliance. Possible attacks that can exploit the lack of integrity protection for \ac{up} data\cite{8835335}, and attacks aiming to de-synchronise the \ac{odu} and \ac{oru}), have been identified. Their impact can be catastrophic, as de-synchronised devices will affect the availability of the system, and missing integrity protection enables the modification of \ac{up} data. While some of those attacks, can be also conducted over the air, the possibility to conduct them in the infrastructure allows scalability to more end devices. Therefore, a practical evaluation is needed for estimating the efforts connected to those attacks. 

For the \ac{ml} part of our upcoming security analysis, a \ac{nrt}, running several xAPPS that are utilizing \ac{ml} to perform  different \ac{ran} functions, is required. The aim is to manipulate the RAN by executing the xAPPS with ill-trained models and monitoring the consequences (potential disruption of operation and malfunction of services). 

\textit{B. Components} For matching those requirements we need a setup includes that includes \ac{ocu}, \ac{odu}, \ac{oru}, and a \ac{nrt}. The selection of these components is based on the total cost of the setup. Therefore, we propose a minimal setup that meets the main requirements of the \ac{oran} architecture. The minimization of cost is a crucial factor for making \ac{oran} available in more universities and research facilities. We present our components selection along with alternatives in Table ~\ref{tab:componentes}. 



\begin{table}[h]
\centering
\caption{Components}
\label{tab:componentes}
\begin{tabular}{@{}lll@{}}
\toprule
Comp & Our Choice     & Alternatives            \\ \midrule
CN         & Free5GC\cite{FREE5GC}        & Open5GS\cite{open5gs}, OAI\cite{OAI_CN}        \\
O-CU       & OAI CU\cite{OAI_RAN}         & AirSpan\cite{airspan}, SRS\cite{srs}     \\
O-DU       & OAI\cite{OAI_RAN}         & AirSpan\cite{airspan}, SRS\cite{srs}     \\
O-RU       & Benetel \cite{BENETEL} & Mavenir\cite{maverin}   \\
RIC     & FLEXRIC\cite{OAI_MOSAIC}      & AirSpan\cite{airspan}, SD-RAN\cite{sdran} \\
UE         & COTS UE (OnePlus)        & Any 5G UE   \\ \bottomrule
\end{tabular}

\end{table}

\noindent \ac{oai}, provides open-source \ac{cu} and \ac{du} units, that are frequently updated to implement many of the O-RAN interfaces. We suggest these units, as \ac{oai} recently demonstrated the Open-Front-Haul implementation. FlexRIC, is a part of the MOSAIC5G Group Project, also an \ac{oai} ongoing project. It serves as a \ac{nrt} and implements the O-RAN E2 interface. For the \ac{cn} functions, we use the distributed version of Free5GC, over a Kubernetes cluster. Free5GC is a \ac{3gpp}-compliant, open-source \ac{cn}. Commercial products used in the setup include a Benetel-manufactured \ac{oru} that supports the 7.2x O-RAN split and a \ac{cots} \ac{ue} that consists of an Android smartphone and a programmable SIM card.
\\
\textit{C. Deployments}

\noindent The project is still in progress and the current setup is in its preliminary stage of development. The \textit{Challenges} section is currently undertaking an analysis to identify the reasons behind this. Two potential deployments have been identified, and their visual representation can be seen in Figure \ref{fig:comp}(a). The initial configuration comprises of a monolithic \ac{gnb} that is linked to the FlexRIC. The controller facilitates  multiple xApps operating concurrently. This particular setup can be utilized to scrutinize the security of the RAN's xApps and \ac{ml}.  The second configuration, on the other hand, is not currently suitable for security analysis purposes, but it will serve as a reference point for the future work and the analysis of the Open-Front-Haul. The future setup, depicted in Figure \ref{fig:comp}(b), will satisfy our two requirements and will allow the \ac{ml} and Architectural-Openness risk analyses.

\section{Challenges}

\textit{A. Incomplete and Undocumented Software}\\ The main challenge we faced during the deployment of our setup, ware the incomplete and undocumented implementation of different solutions offered by SW vendors. Inspecting the code, was very often necessary, to understand the different implemented protocols, or the deployment configurations. Also the current implementation of the different components can be characterised as incomplete. More specifically, FlexRIC lack support for the \ac{cu}\textbackslash \ac{du} split, requiring the use of monolithic gNB deployment, and there is currently no support for the 7.2x interface that enables communication between \ac{ocu} and \ac{odu}. Additionally, there is no support for the O1 and A1 interfaces, making integration of the \ac{smo} and O-Cloud, infeasible.\\\\
\textit{B. Unclear and Non-Interoperable Hardware}\\ The second challenge that we have to deal with, were the not interoperable components offered by \ac{hw} vendors. The same vendors, claiming to offer O-RAN compliant products, often stated compatibility with products only from other specific vendors, with contradictions occurring very often. Unclear specifications, and implementation, was also a major issue related to \ac{hw} elements. To our understanding, many vendors are currently implementing their own flavor of \ac{oran}, as the architecture is evolving rapidly and its specifications are being updated very often.

\section{Future Work}

\begin{figure}[H]
    \centering
    \includegraphics[width=0.4\textwidth]{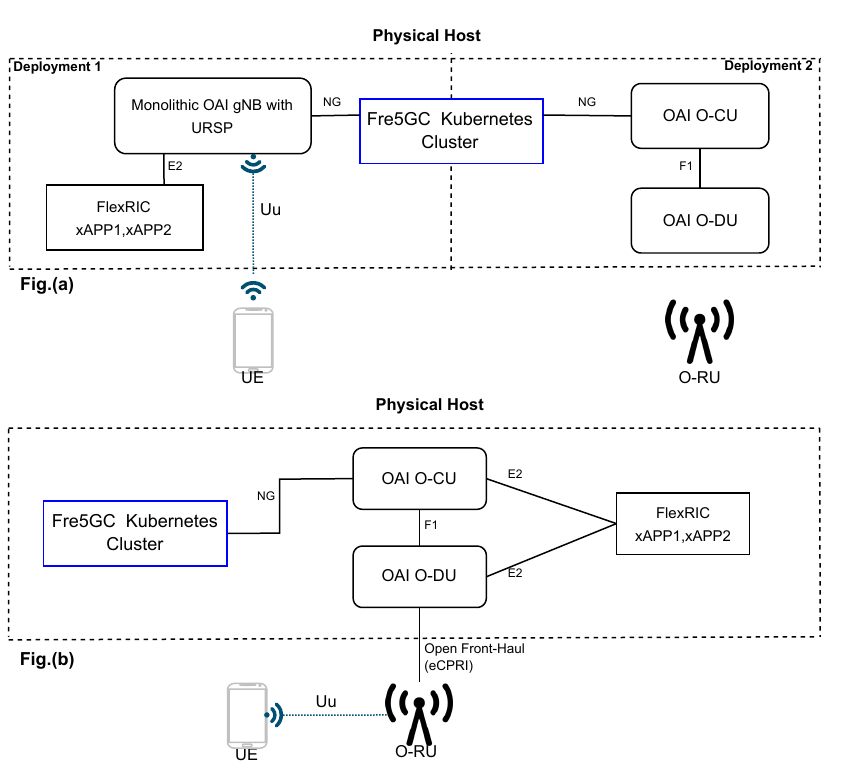}
    \caption{ Current and Future Deployments}
    \label{fig:comp}
    \vspace{-3mm}
\end{figure}

Our primary objective for the future is to merge the two configurations into a single, unified setup. Over the upcoming months, we anticipate the release of a new version from \ac{oai} that will introduce the 7.2x O-RAN split to its components, which will allow for the integration of the \ac{oru} in the current configuration. In regards to the FlexRic, we are also anticipating support for the CU\textbackslash DU split in future updates. However, if this is not achievable, we may consider replacing it with another controller, such as the \ac{ric} utilized in the SD-RAN project. The conflation of the two configurations is represented in Figure \ref{fig:comp}(b).

\noindent This is the most recent O-RAN compliant configuration that can be deployed at a minimal cost. We believe that it is future-proof since the developers of the various components are committed to support and implement every O-RAN interface in the upcoming releases. This will enable the expansion and enhancement of the deployment, eventually with every \ac{oran} element.


\section{Conclusion}
\ac{oran} is an emerging and promising RAN expansion architecture that is popular among the scientific community. Its purpose is to create a smart, scalable, agile, efficient, and vendor-independent \ac{ran}. However, several theoretical risk analyses\cite{5GRAN_ANALYSIS}\cite{MIMRAN2022102890}, have shown that this architecture exposes the \ac{ran} to many new kinds of attacks. The frequent adoption of new specifications has prevented the deployment of any \ac{oran} security test lab. Such a deployment is essential, in-order to enable practical risk analyses of the new architecture, which until now are not-existent. With this poster, we want to present our way to deploying a minimal, future-proof \ac{oran} lab and initiate a discussion with researchers sharing our interests and concerns, and exchange information to achieve our goals and secure \ac{oran}.

\begin{acronym}[ICANN]
    \acro  {oran}   [ORAN]   {Open Radio Access Networks}
    \acro  {ngran} [NG-RAN] {Next Generation Radio Access Networks}
    \acro  {ml}  [ML]   {Machine Learning}
    \acro  {ran}  [RAN]  {Radio Access Networks} 
    \acro{cn}[CN]{Core Network}
    \acro{sw}[SW]{Software}
    \acro{hw}[HW]{Hardware}
    \acro{ue}[UE]{User Equipment}
    \acro{du}[DU]{Distributed Unit}
    \acro{gnb}[gNB]{Next Generation NodeB}
    \acro{cu}[CU]{Central Unit}
    \acro{ric}[RIC]{RAN Intelligent Controller}
    \acro{mwc}[MWC]{Mobile World Congress}
    \acro{bbu}[BBU]{Baseband Unit}
    \acro{ru}[RU]{Radio Unit}
    \acro{odu}[O-DU]{ORAN Radio Unit}
    \acro{ocu}[O-CU]{ORAN Central Unit}
    \acro{oru}[O-RU]{ORAN Distributed Unit}
    \acro{smo}[SMO]{Service Management and Orchestration Framework}
    \acro{oai}[OAI]{Open Air Interface}
    \acro{SRS}[SRS]{Software Radio Systems}
    \acro{3gpp}[3GPP]{3rd Generation Partnership Project }
    \acro{nrt}[nRT RIC]{Near-Real-Time RAN Intelligent Controller}
    \acro{nonrt}[non-RT RIC]{Near-Real-Time RAN Intelligent Controller }
    \acro{cots}[COTS]{Commercial Of The Shelf}
    \acro{ecpri}[eCPRI]{evolved Common Public Radio Interface (eCPRI)}
    \acro{up}[UP]{User Plane}

\end{acronym}

\bibliographystyle{ieeetran}
\bibliography{sample-base}

\end{document}